\documentstyle[11pt,newpasp,twoside,epsf]{article}

\begin{document}

\title{ Euler, Hipparcos and the five dwarfs : \\
        Reconstructing star formation histories in the Local Group}

\author{David Valls--Gabaud$^1$, Xavier Hernandez$^2$ \& Gerry Gilmore$^3$}

\affil{$^1$ Laboratoire d'Astrophysique, UMR CNRS 5572, \\
            Observatoire Midi-Pyr\'en\'ees, 
            14 Av. E. Belin, 31400 Toulouse, France}
\affil{$^2$ Osservatorio Astrofisico di Arcetri,\\
            Largo Enrico Fermi 5, 50125 Firenze, Italy}
\affil{$^3$ Institute of Astronomy, Madingley Road, Cambridge CB3 0HA, UK}

\begin{abstract}
The unprecedented quality of recent colour-magnitude diagrams of
resolved stellar populations in nearby galaxies requires 
state-of-the-art techniques to infer the star formation histories
which gave rise to the observed distributions. We have developed
a maximum likelihood technique, which coupled to a variational
calculus allows us to make a robust, non-parametric reconstruction
of the evolution of the star formation rate. A full Bayesian analysis
is also applied to assess whether the best solutions found are also
good fits to the data. Applying this new method to WFPC2 observations
of five dSph galaxies of the Local Group, we find a wide variety
of star formation histories, with no particular epoch being dominant.
In the case of the solar neighbourhood observed by Hipparcos  we
infer, with an unprecedented resolution of 50 Myr, its star formation
history over the past 3 Gyr. The surprising regularity of star
formation episodes separated by some 0.5 Gyr could be interpreted as
the result of interactions with two spiral arms or the Galactic
bar. These bursts possibly trigger the formation of massive star 
clusters which slowly dissolve into the galactic field.
\end{abstract}


\section{Introduction}
\label{sec_intro}
The history of the star formation rate in galaxies is one of the
main ingredients required to understand their formation and evolution.
Whilst the Roberts' time scale for gas consumption by star formation
gives a rough idea of the evolution of the stellar and gas content, 
more elaborate
models (as summarised, e.g. by Sandage, 1986) can reproduce a wide variety
of observables, from the evolution of the bulge-to-disc ratio in the
Hubble sequence to the integrated colours as a function of redshift. However, in
 sharp contrast with these global views where the star formation rate SFR(t) is
a monotone function of time, current bottom-up hierarchical scenarios of 
structure formation predict far more complicated histories, where star
formation episodes are related to, if not driven by, halo mergers. The
star formation history is, in this context, a more or less faithful 
reproduction of the merger history, rich in both strong and minor events
at redshifts $z < 2$ or so.

While it is possible to understand some of the observed properties of high
redshift galaxies in this framework, a far more direct test can be achieved
with the fossil record of the star formation history itself: the colour-magnitude
diagrams (CMDs) of the resolved stellar populations in nearby galaxies. Progress
in this field has been steady, driven both by ground-based wide-field 
surveys and by deep, high-resolution, narrow-field HST studies, reaching
well below the main sequence turn off point in many satellites of the Local
Group (see Aparicio 1998, Mateo 1998 for recent reviews). 

\section{Objective reconstructions of star formation histories}
\label{sec_theory}
The unprecedented quality of the current CMDs requires powerful tools
to invert them in order to derive the star formation history (SFH) which gave
rise to the observed distribution of stars in these diagrams. The methods
used so far are based on comparisons between the observed CMD and synthetic
CMDs computed assuming a given SFH, and then looking for the best matching
SFH. For instance Tolstoy \& Saha (1996) use a Bayesian likelihood technique,
Mighell (1997) a classical $\chi^2$ optimisation, 
Ng (1998) a Poisson merit function, while Gallart et al. (1999) and
Hurley-Keller et al. (1999) minimise a counts in cells statistic for
well-defined boxes in the CMD. Alternatively, Dolphin (1997) makes a
linear decomposition in terms of fiducial CMDs and solves for the best
matching final CMD, in terms of a   $\chi^2$  statistic.

One of the problems with most of these techniques is that there is no guarantee
that the actual SFH belongs to the set of (usually parametric) SFHs explored
within a family of functions defined a priori. 
For instance, in the case of the Carina dSph, Hurley-Keller et al. (1999) 
find the best fitting 3-burst solution, leading to results which are 
at variance with those from Mighell (1997), who uses a non-parametric approach.
In addition, it is unclear to what extent these best matching solutions
are actually good fits to the observed CMDs. We have therefore developed an entirely
new method based on a combination of Bayesian statistics with variational
calculus which does not suffer from the limitations listed above. Full details of the method are given in Hernandez et al.
(1999, Paper~I), 
 and will not be repeated here.  Very briefly, the method makes 3 key 
assumptions: (1) the metallicity of the ensemble of stars is known and
has a small dispersion; (2) the initial mass function is given and
there are no unresolved binary systems; and (3)
both distance and colour excess are known to within some uncertainties.
The method then 
takes as inputs the positions of $n$ stars in a colour magnitude
diagram, each having a colour $c_i$ and luminosity $l_i$, with
(in this example, uncorrelated) associated errors $\sigma(c_i)$ and $\sigma(l_i)$
respectively.  Using the likelihood technique, we first construct the
probability that the $n$ observed stars resulted from some function
$SFR(t)$. This is given by

\begin{equation}
{\cal L}= \prod_{i=1}^{n} \left( 
\int_{t_0} ^{t_1} \, SFR(t) \, G_{i}(t) \, dt \right),
\label{eq:likelihood}
\end{equation}

\noindent where

\begin{eqnarray}
G_{i}(t)  =  \int_{m_0}^{m_1} {\rho(m;t) \over{2 \pi \sigma(l_i)
 \sigma(c_i)}} \,  
 \exp\left(-D(l_{i};t,m)^2 \over {2 \sigma^2(l_i)} \right) 
\exp\left(-D(c_{i};t,m)^2 \over {2 \sigma^2(c_i)} \right) \; dm 
\end{eqnarray}

In this expression $\rho$ is the density of stars of mass $m$ along the isochrone of
age $t$, and only depends on the assumed IMF and the set of stellar 
tracks (and in particular the durations of the different evolutionary phases).
The $D$ factors are the differences in luminosity and colour of
the observed star $i$ with respect to the luminosity and colour of a
star of mass $m$ at time $t$. We refer to $G_i(t)$ as the likelihood
matrix, since each element represents the probability that a given
star $i$ was actually formed at time $t$ with any mass.

Following the discussion of Paper~I, we may write the condition that
the likelihood has an extremal as the variation $\delta {\cal L}(SFR)
= 0 $, allowing a full variational calculus analysis to be used.
Developing first the product over $i$ using the chain rule,
and dividing the resulting sum by ${\cal L}$, one obtains

\begin{equation}
\sum_{i=1}^{n} \left(
{\delta \int_{t_0} ^{t_1} SFR(t) \, G_{i}(t) \, dt} \over {\int_{t_0}^{t_1}
 SFR(t) \, G_{i}(t) \, dt} \right) =0
\label{eq:sum}
\end{equation}

\noindent Introducing the new variable $Y(t)$ defined as

\begin{equation}
Y(t)=\int{ \sqrt {SFR(t)} \, dt}\;  \Longrightarrow \;  SFR(t)=\left( {dY(t)
\over dt} \right)^2
\end{equation}

\noindent
into Equation~\ref{eq:sum} we can develop the Euler equation to yield 

\begin{equation}
{d^2 Y(t)\over dt^2}\sum_{i=1}^{n} \left( G_{i}(t) \over I(i)\right)
=-{dY(t)\over dt}\sum_{i=1}^{n} \left( dG_{i}/dt \over I(i)\right)
\end{equation}

\noindent where 
\begin{equation}
I(i)=\int_{t_0}^{t_1} SFR(t)\,  G_{i}(t) \, dt
\end{equation}
is an integral constraint.

We have now transformed what was an optimisation problem, finding the
function that maximises the product of integrals defined by
Equation~\ref{eq:likelihood}, into an integro-differential equation with a boundary
condition (at either $t_o$ or $t_1$) which can be solved by iteration to
get, non-parametrically, the function $SFR(t)$. 
Further details on the numerical aspects of the procedure are
available in Paper~I. It is important to point out that our method
has distinctive advantages over other techniques: (1) the variational
calculus allows a fully non-parametric reconstruction, free of any
astrophysical preconceptions; (2) there is no time-consuming 
comparisons between CMDs, since the function (not the parameters) $SFR(t)$
that maximises the likelihood if solved for directly; (3) the CPU
scales linearly with the time resolution required in the reconstruction.

Paper~I also presents a detailed study on the influence of
the assumptions that were made: the effect
of changing the IMF or the presence of unresolved binaries is essentially
a normalisation problem, which does not change the overall shape or
localisation of a burst of star formation, while a wrong estimate of the metallicity has
drastic effects on the position of a burst, a result of the age-metallicity
degeneracy. See also Paper~I for the effect of photometric
uncertainties in the effective time resolution of the reconstructed SFH. 
Note that since the IMF and the fraction of binaries (and the distribution
function of their mass ratios) are unlikely to be measured, an absolute
normalisation of the star formation rate cannot be achieved.

\begin{figure}[th]
\plotfiddle{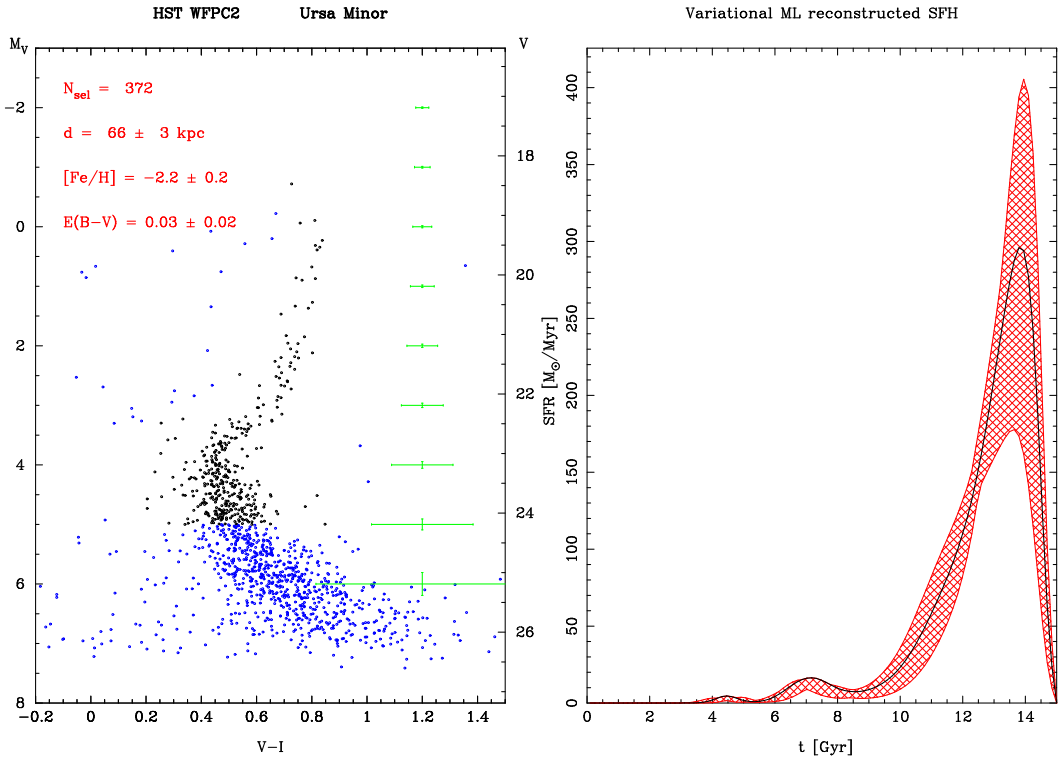}{9.8cm}{-90.0}{55.}{55.}{-214}{324}
\caption{Colour-magnitude diagram and  SFR history of UMi.}
\protect\label{fig_umi}
\end{figure}

\section{Star formation histories of Local Group dSph satellites}
\label{sec_dsph}
Given the main assumption made so far in order to derive a fully non-parametric
variational reconstruction of the SFH, namely a determination of the
metallicity of the ensemble of stars, only a few galaxies fulfill this
requirement. In addition, in order to minimise any possible systematic
effects between different data sets and reduction procedures, as well as
crowding/blending effects, we selected
from the HST archive an homogeneous sample of five dSph galaxies (Leo I,
Leo II, Draco, Carina and Ursa Minor) of the Local
Group which were reduced with the same procedures. It is only such an internally
consistent data set which allows robust comparisons between different
galaxies, with the proviso that the volumes sampled by the WFPC2 field may
not be representative of the entire galaxy. 
Full details of the SFH reconstructions for these galaxies are presented
in Hernandez et al. (2000, Paper~II).

The dSph UMi can be used as an example of our procedure. Figure~\ref{fig_umi}
gives the WFPC2 CMD we used as input. Note the increase of the 
2$\sigma$ error bars with decreasing luminosity, and the removal of stars on the
horizontal branch (incompatible
with the evolutionary phases included in our isochrones). Since 
incompleteness may be important, and isochrones are
degenerate at this level, we also remove stars below $M_V = 5$, so we
are left with $N=372$ stars in total. For the fiducial values of
distance, metallicity and colour excess indicated (slightly different
but consistent with the more recent determinations by Mighell 
\& Burke, 1999), the variational
method gives the reconstructed SFH indicated with the black line on the
right panel of Fig.~\ref{fig_umi}. This function maximises the
likelihood, as defined in Eq.~\ref{eq:likelihood}, for the given data
set. To assess the robustness of this reconstruction, many
functions were reconstructed by changing the observational
parameters within their error bounds indicated on Fig.~\ref{fig_umi}.
The {\sl envelope} of such solutions is given as the shaded area on the
figure, and represents the uncertainties in the reconstruction given
by the uncertainties in the observations, so that {\sl any} function
that will be contained within these bounds will maximise ${\cal L}$,
given these errors. The bulk of the stellar population in UMi was
therefore formed more than 12 Gyr ago, with a peak at 14 Gyr. Some
residual activity may have been present at younger ages, but it is
caused by stars bluer than the main turn off point or by
blue stragglers.  Given the uncertainties
in the photometry, we cannot resolve any episodes in the star formation
rate within the main burst at 14 Gyr, but we can conclude that the burst
lasted less than 2 Gyr (FWHM).

Having found the best solution (in terms of maximising the likelihood)
does not guarantee that it is also a good solution. The classical way
to assess the goodness of fit, via Fisher's information matrix, does
not apply here (there are no parameters), so we apply again Bayesian
techniques to check whether the observed CMD could result from the CMDs
produced by the best reconstruction of the SFH. To do this, we do a
counts in cells analysis and apply
Saha's W statistic (Saha, 1998), basically
\begin{equation}
W \; \propto \; \prod_{i=1}^{B} \frac{ (m_i+n_i)! }{ m_i!  \; n_i!  }
\end{equation}
with $B$ number of cells the CMD is divided into, and $m_i$ and $n_i$ 
the number of stars in cell $i$ in two CMDs. To compute the distribution
function of $W$, a series of model-model comparisons are made, that is, synthetic
CMDs resulting from realisations of the reconstructed SFH are compared pairwise. To check 
whether the observed CMD could arise from this distribution, the $W$
statistic resulting from comparisons between the observed CMD and a
series of CMD realisations are made. If the distribution functions
are compatible, at some statistical level, the hypothesis that the
observed CMD can be produced by the reconstructed SFH is accepted.

In the case of UMi, the mean and dispersion of the model-model $W$
distribution is 47 $\pm$ 5, while the data-model distribution gives
44 $\pm$ 4, so we can deduce that the synthetic CMDs are compatible at
better than 1$\sigma$ with the observed CMD. The best solution is also
a good solution.

\begin{figure}[th]
\plotfiddle{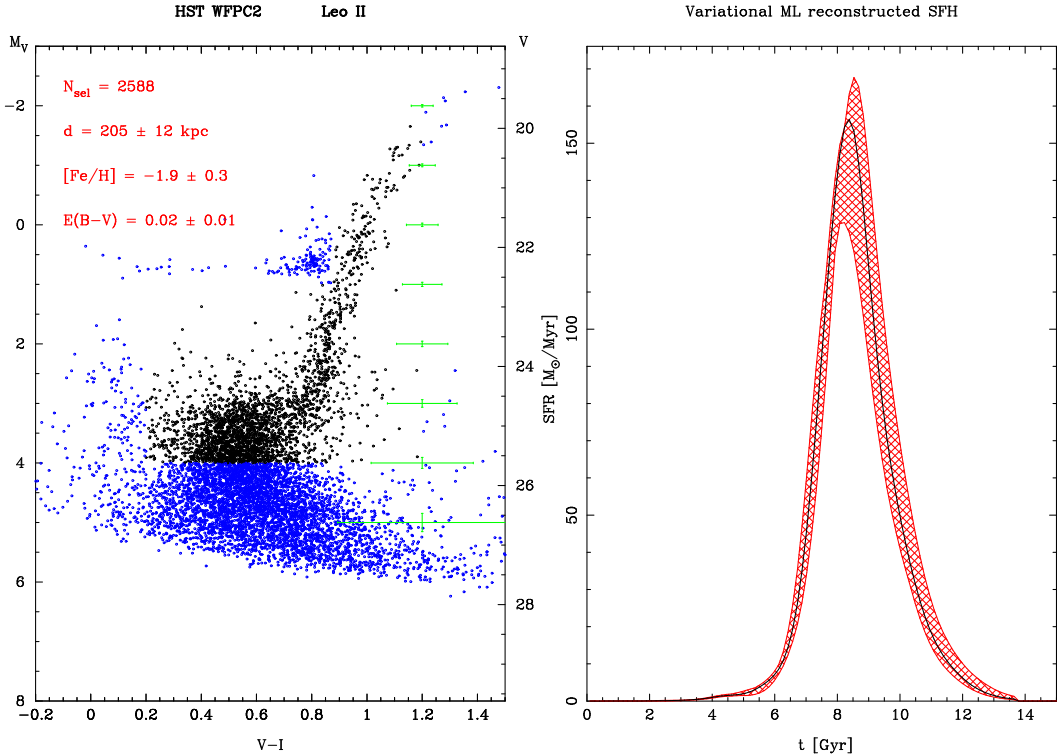}{9.8cm}{-90.0}{55.}{55.}{-214}{324}
\caption{Colour-magnitude diagram and  SFR history of Leo II.}
\protect\label{fig_leo2}
\end{figure}

The dSph Leo~II (see Fig.~\ref{fig_leo2}) presents a similar case, whereby
the bulk of stars was formed in a burst slowly rising at 12 Gyr with
a maximum around 8 Gyr and an abrupt decrease at 6 Gyr. However in this
case, the very sensitive $W$ statistic says that the observed CMD is
not compatible (at the 2$\sigma$ level) with the synthetic diagrams
resulting from this best solution. The reason is likely to be the relatively
large spread in metallicity, around 0.3 dex, showing the limitation of
the method.


\begin{figure}[th]
\plotfiddle{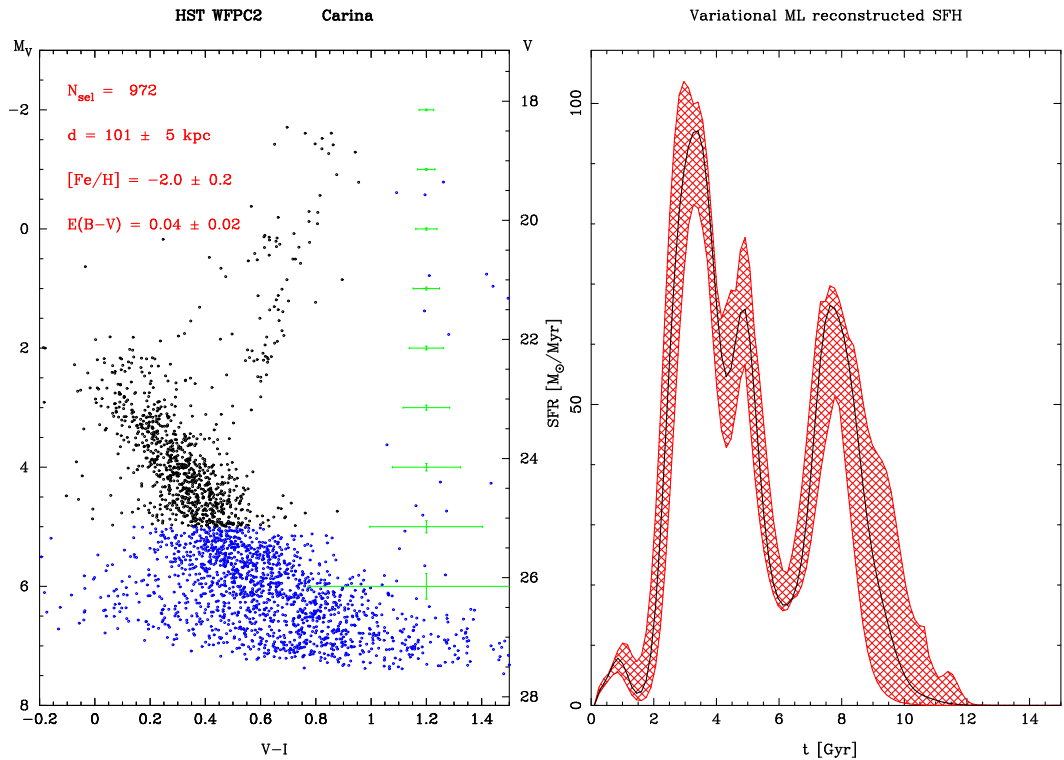}{9.8cm}{-90.0}{55.}{55.}{-214}{324}
\caption{Colour-magnitude diagram and SFR history of Carina.}
\protect\label{fig_carina}
\end{figure}

The case of the Carina dSph (Fig.~\ref{fig_carina}) is more interesting.
There are clearly two sub-giant branches, identified with two epochs of
star formation at 8 and 3 Gyr. Given the 2$\sigma$ errors indicated in
Fig.~\ref{fig_carina} and simulated CMDs with these errors, we can say
that the bursting episodes in Carina lasted more than a 1 Gyr, and that
there was no epoch in the 2--8 Gyr interval without star formation
activity. The $W$ statistic indicates a good solution. It is however
puzzling that such an extended star formation history, over 6 Gyr, 
resulted in such a small metallicity, with a dispersion of 0.2 dex or less.
Either these values are only representative of the tip of the giant
branch, or the enrichment in metals had a non standard history.

\begin{figure}[th]
\plotfiddle{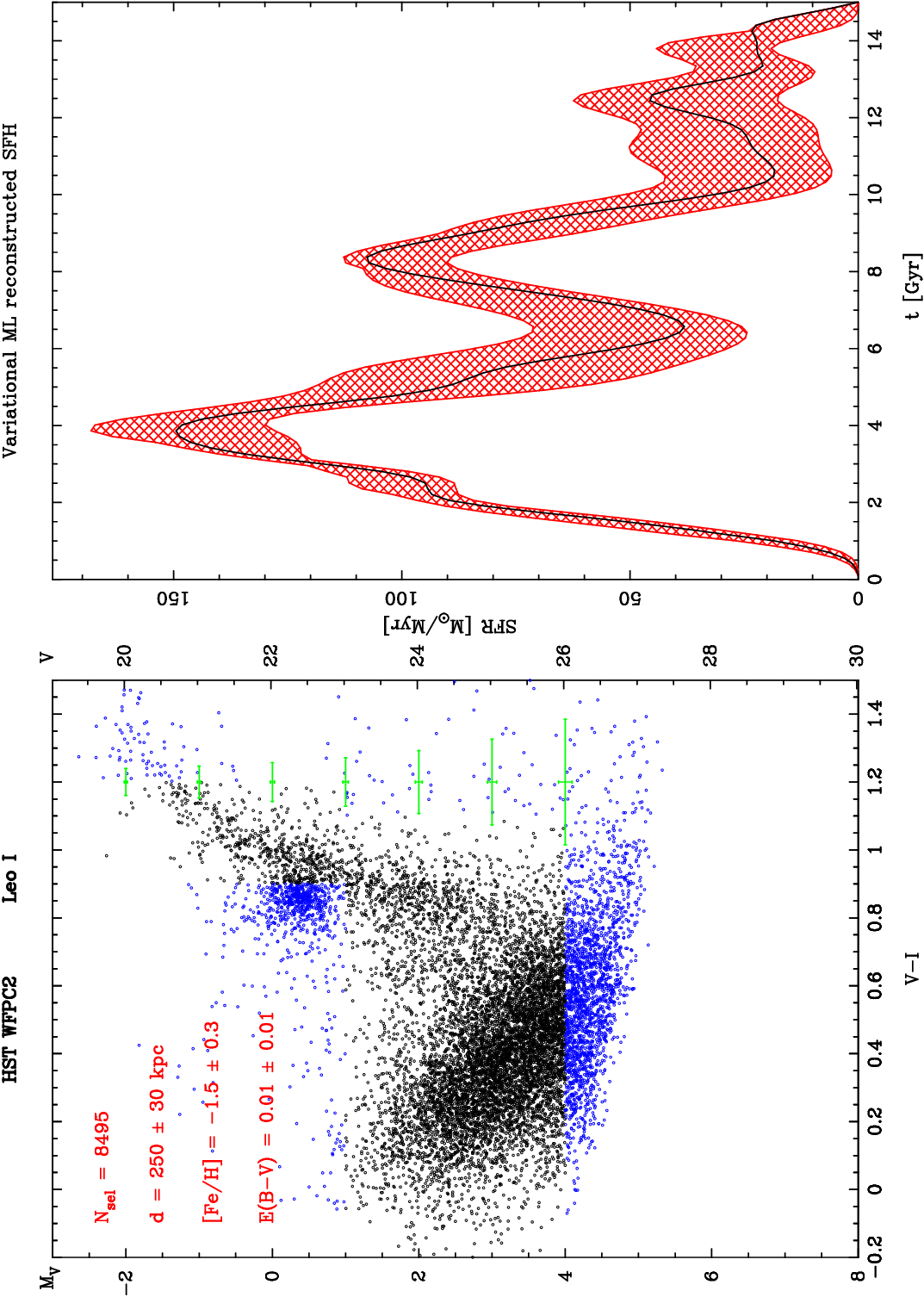}{9.8cm}{-90.0}{55.}{55.}{-214}{324}
\caption{Colour-magnitude diagram and  SFR history of Leo I.}
\protect\label{fig_leo1}
\end{figure}

Leo~I presents a similar, extended, history of star formation (see Fig.~\ref{fig_leo1}). Some activity was present around 10--13 Gyr ago,
compatible with the presence of horizontal branch stars recently
detected at the NTT (Held et al. 2000) but not present in our HST data.
The bulk of the observed stellar populations was formed  continuously,
with peaks of activity at 8 and 4 Gyr, in agreement with an
independent analysis by Gallart et al. (1999). In this case however,
the $W$ statistic indicates that the maximum likelihood solution is not
a good solution at the 2$\sigma$ level. Again, a likely explanation is
that the dispersion in metallicity (around 0.3 dex) is too large for the
method to work. A full bi-variate non-parametric reconstruction of both
$SFR(t)$ and $Z(t)$ is required.

If the Local Group is a representative volume of the Universe, the
star formation activity at high redshift was dominated by these dSph
that dominate the luminosity function. A properly averaged SFH does
not show any significant epoch in the comoving density of SFR (Tolstoy 
1999, Paper~II). There seems to be no link either between the epoch of the
bursts and perigalacticon passages of these satellites. 

\section{The evolution of the SFR in the solar neighbourhood}
\label{sec_solar}

The evolution of the star formation rate in the Galactic disc is another
basic function required to understand the formation of the disc, its
chemical evolution and, more generally, the luminosity evolution of
spiral galaxies. Previous attempts at determining the SFH in the
Galaxy have relied on indirect methods, basically relations between
age and some astrophysical property --such as chromospheric activity
or metallicity-- and then correct with evolutionary models for
the stars which have disappeared from the sample. A good example
of the complexities of these techniques is provided by Rocha-Pinto
et al. (2000b). Here we use the method outlined in \S~\ref{sec_theory} 
since the solar neighbourhood has a small dispersion in metallicity 
centred on the solar value. Further details will be found in Hernandez
et al. (2000b, Paper~III).

\begin{figure}
\begin{minipage}{0.9\textwidth}
\plotfiddle{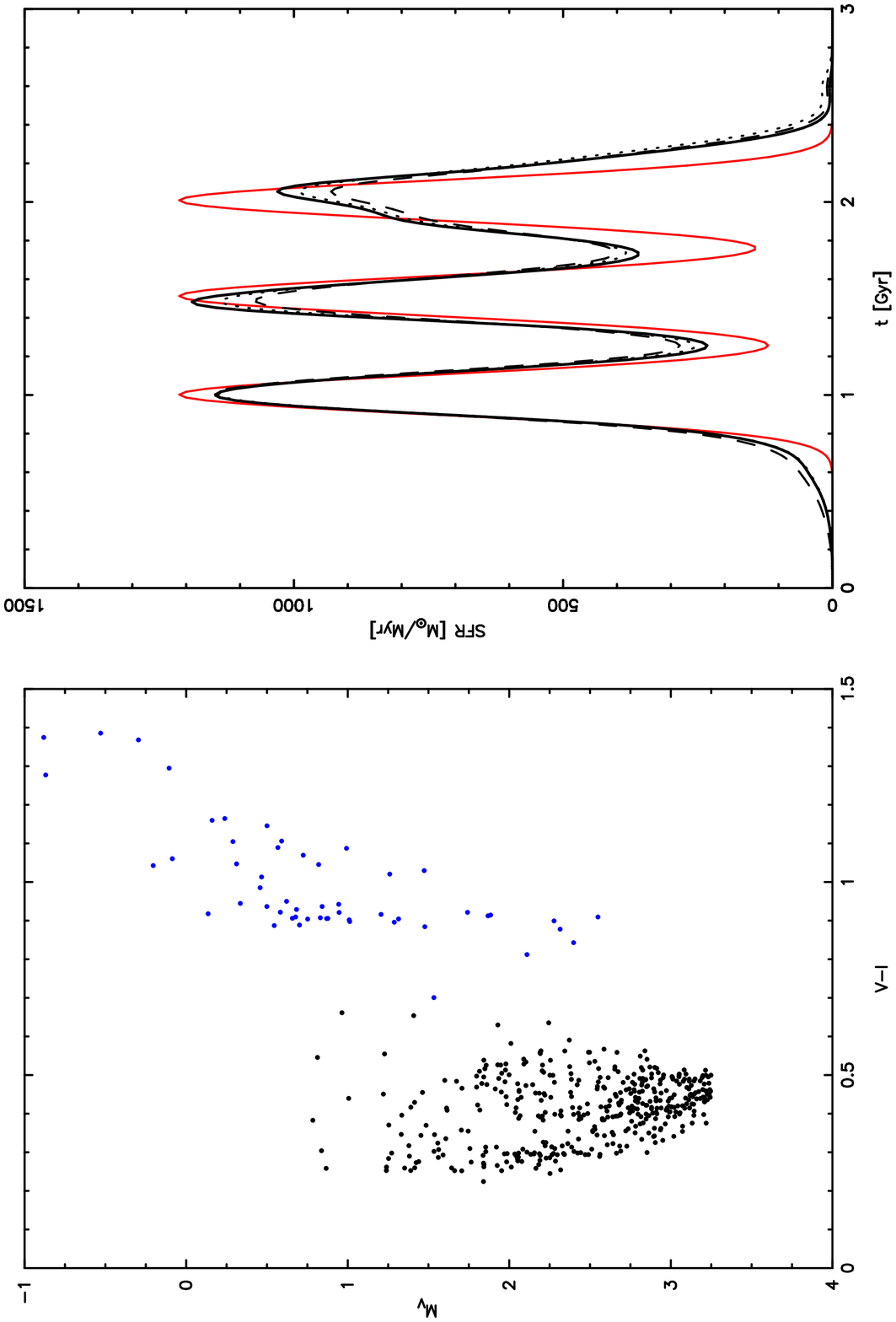}{7.8cm}{-90.0}{55.}{55.}{-214}{324}
\vskip 13mm
\plotfiddle{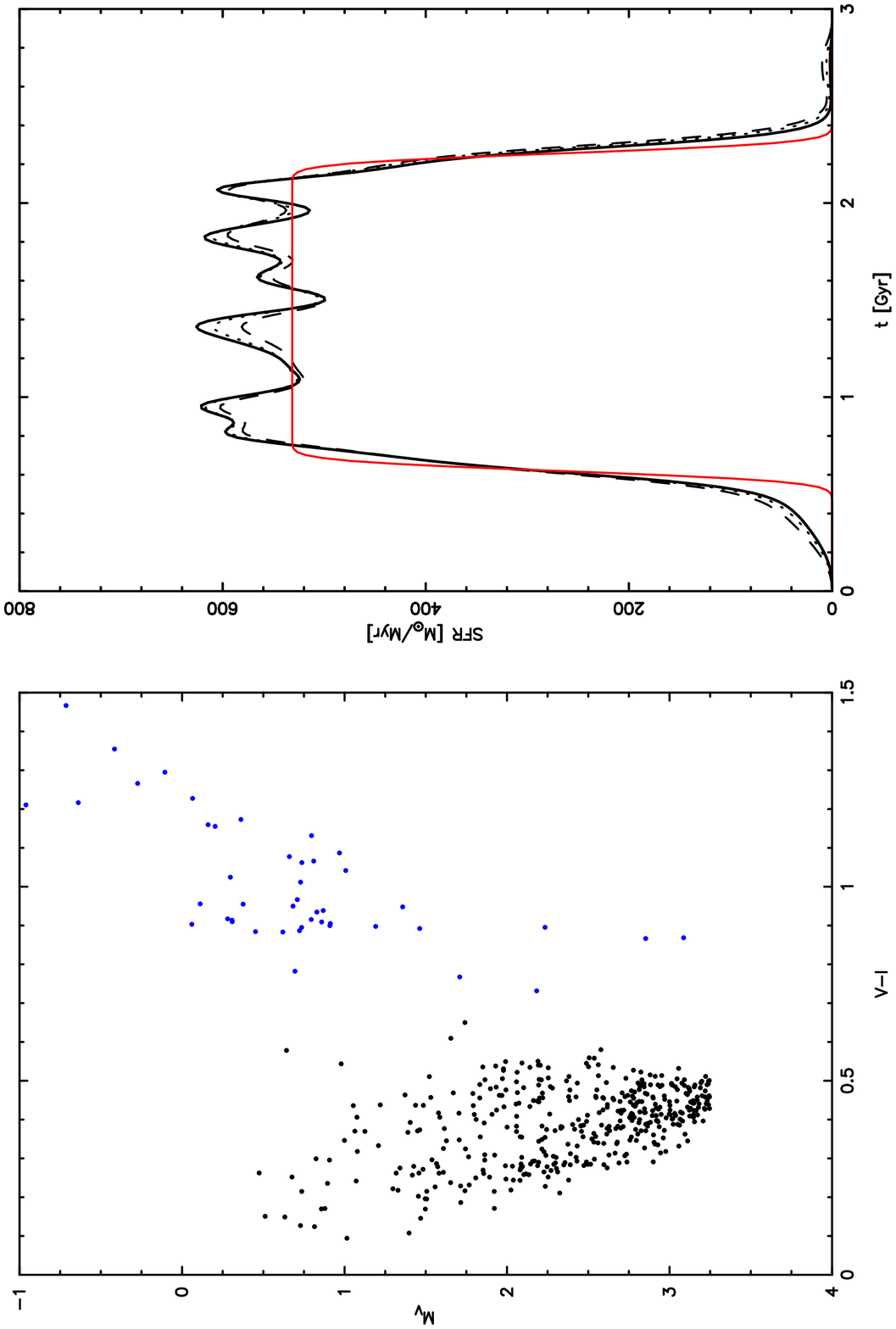}{7.8cm}{-90.0}{55.}{55.}{-214}{324}
\end{minipage} \hfill
\caption{Testing the inversion technique for Hipparcos realisations.}
\protect\label{fig_hip_tests}
\end{figure}

The first step is to define a volume-limited sample of well-measured
stars, and the ideal catalogue for this task is obviously the Hipparcos
catalogue. This provides a direct way to infer the SFH of the solar
neighbourhood, as opposed to more indirect methods (e.g. Rocha-Pinto
et al. 2000a). Although the completeness of Hipparcos varies both with
spectral type and galactic latitude, a cut at $V=7.9$ for the sample
with parallax errors smaller than 20\% provides a reasonable sub-sample,
once binaries and variable stars are removed. 
A typical cut at $V=7.25$ can produce an absolute-magnitude limited
sample complete in volume with a well understood error distribution. The
absolute magnitude limit, say at $M_V=3.15$, implies that only stars
younger than about 3 Gyr enter the sample, but the kinematical and geometrical
corrections are minimised. With such samples, and their error distributions,
we simulated synthetic CMDs with two different SFHs: a 3 burst scenario
and a constant SFR over 1.5 Gyr. Figure~\ref{fig_hip_tests} shows that
our maximum likelihood variational method reconstructs correctly such
star formation histories, even though the number of stars is very small. 
Note also that the CMDs look very similar, yet they are produced by 
very different SFHs.

\begin{figure}
\plotfiddle{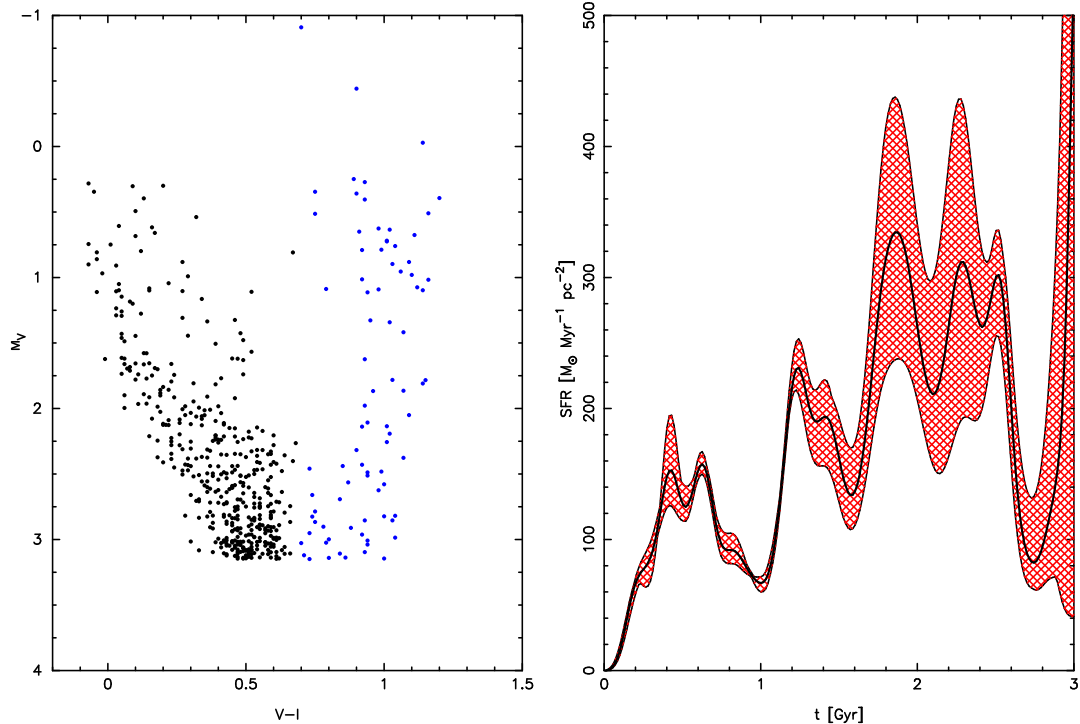}{9.2cm}{-90.0}{55.}{55.}{-214}{324}
\caption{Colour-magnitude diagram and SFH of the solar neighbourhood
subsample observed by HIPPARCOS. Only stars bluer than $V-I=0.7$ 
are considered in the reconstruction. }
\protect\label{fig_hip_sfh}
\end{figure}

The inversion procedure applied to the actual Hipparcos diagram for
the solar neighbourhood is shown on Fig.~\ref{fig_hip_sfh}. The superb
quality of the data allows us to reconstruct its SFH with the
unprecedented resolution of 50 Myr, and a
clear pattern emerges: over a roughly low-level constant SFR, there are several
distinct peaks, separated by about 0.5 Gyr. As before, the envelope
gives the area where any function will maximise the likelihood, when
different $M_V$ cuts are applied. Note that the envelope increases
with age, a reflection of larger uncertainty. This best solution is also
a good solution 
(see Paper~III). 

 A possible interpretation of the quasi-periodic episodes in the
star formation rate of the solar neighbourhood may be given in terms
of interactions with either spiral arms or the galactic bar.
 As the pattern speed and the circular velocity are in
general different, the solar neighbourhood periodically crosses an arm
region, where the increased local gravitational potential might
possibly trigger an episode of star formation. In this case, the time
interval $\Delta t$ between encounters with an arm at the solar
neighbourhood is

\begin{equation}
\Delta t \; = \; \frac{0.22 \, {\rm Gyr}}{m} \; \left(\frac{\Omega}{29 \,
 {\rm km \, s}^{-1} {\rm kpc}^{-1}}\right)^{-1} \; 
 \left| \frac{\Omega_p}{\Omega} - 1  \right|^{-1}
\end{equation}
 
\noindent
where $m$ is the number of arms in the spiral pattern. 
Recent determinations tend to point to large
values for the pattern speed 
 (e.g.  Mishurov et al. 1979, Avedisova 1989, Amaral and L\'epine 1997) 
 close to $\Omega_p \sim 23 - 24$ km
s$^{-1}$ kpc$^{-1}$, which would imply that the regularity
present in the reconstructed $SFR(t)$ would be consistent with a
scenario where the interaction of the solar neighbourhood with a
two-armed spiral pattern would have induced the star formation
episodes we detect. This is consistent with the current knowledge on
the nearby spiral structure (Vall\'ee 1995), and is reminiscent of the explanations put
forward to account for the inhomogeneities observed in the Hipparcos
velocity distribution function, where well-defined branches associated
with moving groups of different ages (Chereul et al. 1999, Skuljan et
al. 1999, Asiain et al. 1999) could perhaps be also associated with an
interaction with spiral arm(s), although in this case the time scales
are much smaller.  Of course, other explanations are possible; for
example the cloud formation, collision and stellar feedback models of
Vazquez \& Scalo (1989) predict a phase of oscillatory star formation
rate behaviour as a result of a self-regulated star formation
r\'egime. Close encounters with the Magellanic Clouds have also been
suggested to explain the intermittent nature of the star formation
rate, though on longer time scales (Rocha-Pinto et al. 2000b).

\section{Conclusions}

We put forward 
a maximum likelihood technique, coupled to a variational
calculus, which allows the robust, non-parametric reconstruction
of the evolution of the star formation rate from the information
contained in colour-magnitude diagrams. A full Bayesian analysis
is also applied to assess whether the best solutions found are also
good fits to the data. Its main limitation, at the moment, is the 
prior knowledge of the metallicity of the ensemble of stars in a CMD.

For the first time, an objective reconstruction of star formation
histories without any a priori or model-dependent information is applied
to an homogeneous sample of dwarf spheroidals of the Local Group.
We find a wide variety of SFHs, with bursts of activity uncorrelated
to any special epoch or event, like perigalacticon passages. 
Among many other things, this also implies that
late accretion was not important in the formation of the Galactic halo
(Gilmore et al. 2000). 

In the solar neighbourhood observed by Hipparcos,  we
infer --with an unprecedented resolution of 50 Myr-- its star formation
history over the past 3 Gyr, finding a surprising regularity of star
formation episodes separated by some 0.5 Gyr. A possible
explanation  is that the solar neighbourhood interacted 
 with two spiral arms or the Galactic
bar,  triggering star formation at each interaction. 
These bursts are likely to induce the formation of massive star 
clusters which slowly dissolve into the galactic disc.


\begin{references}


\reference Amaral L.H., L\'epine J.R.D., 1997, MNRAS, 286, 885
\reference Aparicio A., 1998, in IAU Symp 192, The stellar content of
Local Group galaxies, ed. P. Whitelock \& R. Cannon (San Francisco: ASP), 20
\reference Asiain R., Figueras F., Torra J., 1999, A\&A, 350, 434
\reference Avedisova V.S., 1989, Astrophys. 30, 83
\reference Chereul E., Cr\'ez\'e M., Bienaym\'e O., 1999, A\&A Suppl. 135, 5
\reference Dolphin, A., 1997, New Ast., 2, 397
\reference Gallart C., Freedman W., Aparicio A., Bertelli G., Chiosi C., 1999, AJ, 118, 2245
\reference Gilmore G., Hernandez X., Valls-Gabaud D., 2000, The Galactic halo:
from globular clusters to field stars, 35th Li\`ege Conf., ed. A. Noels et al.
 in press (astro-ph/{\tt 9910409})
\reference Held, E.V. et al., 2000, ApJ, 530, L85
\reference Hernandez X., Valls-Gabaud D., Gilmore G., 1999, MNRAS, 304, 705 (Paper~I) 
\reference Hernandez X., Gilmore G., Valls-Gabaud D., 2000a, MNRAS, 317, 83 (Paper~II) 
\reference Hernandez X., Valls-Gabaud D., Gilmore G., 2000b, MNRAS, 316, 605 (Paper~III) 
\reference Hurley-Keller D., Mateo M., Nemec J., 1998, AJ, 115, 1840
\reference Mateo M., 1998, ARA\&A, 36, 435
\reference Mighell K.J., 1997, AJ, 114, 1458
\reference Mighell K.J., Burke, C.J., 1999, AJ, 118, 366
\reference Mishurov Y.N., Pavloskaya E.D., Suchkov A.A., 1979, AZh, 56, 268
\reference Ng, Y.K., 1998, A\&AS, 132, 133
\reference Rocha-Pinto H.J., Scalo J., Maciel W.J., Flynn C., 2000a, ApJ, 531, L115
\reference Rocha-Pinto H.J., Maciel, W.J., Scalo J.,  Flynn C., 2000b, A\&A, submitted (astro-ph/{\tt 0001383})
\reference Saha P., 1998, AJ, 115, 1206
\reference Sandage, A. 1986, A\&A, 161, 89
\reference Skuljan, J., Hearnshaw, J.B., Cottrell, P.L., 1999, MNRAS, 308, 731
\reference Tolstoy E., 1999, in Dwarf galaxies and cosmology, ed. T.X. Thuan, C. Balkowski, V. Cayatte, J. Tran Thanh Van (Paris: Editions Fronti\`eres)
\reference Tolstoy E., Saha A., 1996, ApJ, 462, 672
\reference Vall\'ee, J.P., 1995, ApJ, 454, 119
\reference Vazquez E.C., Scalo J.M., 1989, ApJ, 343, 644 


\end{references}
\end{document}